\documentclass[a4paper]{article}
\usepackage{spconf,amsmath,graphicx,amssymb,CJK, indentfirst,amsmath,multirow,balance,appendix,booktabs,setspace}
\usepackage[colorlinks,linkcolor=black]{hyperref}
\usepackage{INTERSPEECH2020}
\title{IEEE SLT 2021 Alpha-mini Speech Challenge: Open Datasets, Tracks, Rules and Baselines}
\name{Yihui Fu\textsuperscript{1,*},
      Zhuoyuan Yao\textsuperscript{1,*}\footnote{* Contribute equally},
      Weipeng He\textsuperscript{2},
      Jian Wu\textsuperscript{1},
      Xiong Wang\textsuperscript{1},
      Zhanheng Yang\textsuperscript{1},
      Shimin Zhang\textsuperscript{1},
      Lei Xie\textsuperscript{1,**}\footnote{** Corresponding author},
      Dongyan Huang\textsuperscript{3},
      Hui Bu\textsuperscript{4},
      Petr Motlicek\textsuperscript{2},
      Jean-Marc Odobez\textsuperscript{2}}

\address{\textsuperscript{1}Audio, Speech and Language Processing Group (ASLP), School of Computer Science, \\
      Northwestern Polytechnical University, Xi'an, China \\
      \textsuperscript{2}Idiap Research Institute, Switzerland\\
      \textsuperscript{3}UBTECH Technology Co., Ltd. Shenzhen, China\\
      \textsuperscript{4}AISHELL Technology Co., Ltd. Beijing, China
}
\email{}
\begin{document}
  %
\maketitle
\begin{abstract}
The IEEE Spoken Language Technology Workshop (SLT) 2021 Alpha-mini Speech Challenge (ASC) is intended to improve research on keyword spotting (KWS) and sound source location (SSL) on humanoid robots. Many publications report significant improvements in deep learning based KWS and SSL on open source datasets in recent years.  For deep learning model training, it is necessary to expand the data coverage to improve the model robustness. Thus, simulating multi-channel noisy and reverberant data from single-channel speech, noise, echo and room impulsive response (RIR) is widely adopted. However, this approach may generate mismatch between simulated data and recorded data in real application scenarios, especially echo data. In this challenge, we open source a sizable speech, keyword, echo and noise corpus for promoting data-driven methods, particularly deep-learning approaches on KWS and SSL. We also choose Alpha-mini, a humanoid robot produced by UBTECH equipped with a built-in four-microphone array on its head, to record development and evaluation sets under the actual Alpha-mini robot application scenario, including environmental noise as well as echo and mechanical noise generated by the robot itself for model evaluation. Furthermore, we illustrate the rules, evaluation methods and baselines for researchers to quickly assess their achievements and optimize their models.
\end{abstract}
\section{Introduction}
\label{sec:intro}
Robots, as useful assistants and playmates, are becoming more and more popular in people's daily life. As the first chain of human-robot speech interaction (HRSI), the accuracy and efficiency of speech interaction have an important impact on the interaction effectiveness and user experience. In typical HRSI scenarios, robot always work in a very complex acoustic scene, including users' voices, background noise, and the voice of the robot itself (echo and mechanical noise). To activate the speech interactions between users and devices, a standby keyword spotting (KWS) module, also known as wake-up word detection, is particularly important to detect predefined keyword in the audio stream to trigger voice interactions. A good KWS system needs to maintain high robustness with low false rejections and false alarms under the constraint of low computation cost. Meanwhile, accurate sound source location (SSL) can provide essential cues for subsequent beamforming, speech enhancement and speech recognition algorithms. In home environments, the following interferences pose great challenges to HRSI: 1) various types of noises from TV, radio, other electrical appliances and human talking, 2) echoes from the loudspeaker(s) equipped on the robot, 3) room reverberation and 4) noises from the mechanical movements of the robot. These noise interferences complicate KWS and SSL to a great extent. Thus, robust algorithms are highly in demand. \par

Conventional KWS system has been developed maturely, including large vocabulary continuous speech recognition (LVCSR) based lattice search~\cite{LVCSR, LVCSR2, LVCSR3}, hidden Markov model (HMM) based keyword-filler method~\cite{HMM,HMM2,Choisy2007Dynamic}, discriminative models based on large-margin formulation or recurrent networks and query-by-example (QbyE) based template matching approaches~\cite{Hou2017Investigating,Chen2015Query,hou2020mining,yuan2019verifying,yuan2020fast}. Recently, with the development of deep learning and its successful applications, deep KWS frameworks have been introduced~\cite{DNN, DNN2, Retsinas2018Exploring, fernandez2007application,higuchi2020stacked,adya2020hybrid, wang2019virtual}. In the deep KWS family, an acoustic model is trained to predict the sub-word of keyword and a posterior handling method is followed to generate a confidence score of the whole keyword. These approaches are highly attractive to deploy on edge-device with small footprint and low latency, as the size of the model can be easily controlled and no complicated graph-search is involved. Besides, attention-based end-to-end method has also been introduced to the KWS task~\cite{Shan2018} and further performance improvement has been observed which significantly simplifies the model structure and the decoding process. Another trick to boost KWS performance recently is to employ a two-stage strategy, where a first-stage detector provides candidates to the second stage to make the final decision~\cite{wu2018monophone, sigtia2020multi}. \par

SSL has been studied for decades. Conventionally, generalized cross correlation with phase transform (GCC-PHAT)~\cite{knapp1976generalized}, steered-response power with phase transform (SRP-PHAT)~\cite{knapp1976generalized} and multiple signal classification (MUSIC)~\cite{Schmidt1986Multiple} are among the most popular approaches. These traditional signal processing based methods are analytically derived with the assumptions about the signal, noise and environment such as the noise is white and the SNR is higher than 0dB, etc. Recently, Lin et al. investigated the reverberation-robust localization approach of using redundant information of multiple microphone pairs and proposed the OnsetMCCC and MCC-PHAT methods~\cite{Lin2018Reverberation,Lin2018JOINTLY}. With the rapid development of deep learning based speech enhancement and separation, several methods were shown to achieve promising performance on the SSL task. In ~\cite{pertila2017robust,Xu2017Weighted}, the authors estimated the masks of target speech to improve the robustness of conventional cross-correlation-based, beamforming-based and subspace-based algorithms for SSL estimation in environments with strong noise and reverberation. In ~\cite{pertila2017robust,Wang2018Robust}, the authors utilized the ideal ratio mask (IRM) and its variants and considered direct sound as the target signal, which leads to high localization accuracy.\par

Although many approaches have addressed the problem of KWS and SSL, there have been only a few studies evaluate the ability of KWS and SSL on humanoid robots with challenging acoustic conditions. On the other hand, large-scale dataset on robot for KWS or SSL is still extremely deficient. He et al. proposed multiple speaker detection and localization dataset recorded by Pepper robot ~\cite{he2018deep}. Lollmann et al. published acoustic source localization and tracking dataset on LOCATA challenge ~\cite{Lollmann2018The}. However, these datasets neither consider the scenario of echo nor have good coverage of different room sizes and reverberation scenarios. Thus it is necessary to release a sizable dataset for KWS and SSL based on humanoid robot and a common platform to better tackle the problem of HSRI in real application scenarios.\par

In this paper, we address the necessity of solving the problem of KWS and SSL on humanoid robot in noisy and echo scenario. It is expected that researchers from both academia and industry can promote the problem solving through this challenge. The rest of the paper is organized as follows. In Section 2, we give detailed introduction of dataset to release.  In Section 3 and Section 4, the details of rules, evaluation method and baselines of KWS and SSL tracks are introduced. Other information about participating the challenge is available in Section 5. A conclusion is drawn in Section 6.
\section{Datasets}
\label{sec:format}
\begin{figure}[htb]
    \centering
    \centerline{\includegraphics[width=6.0cm]{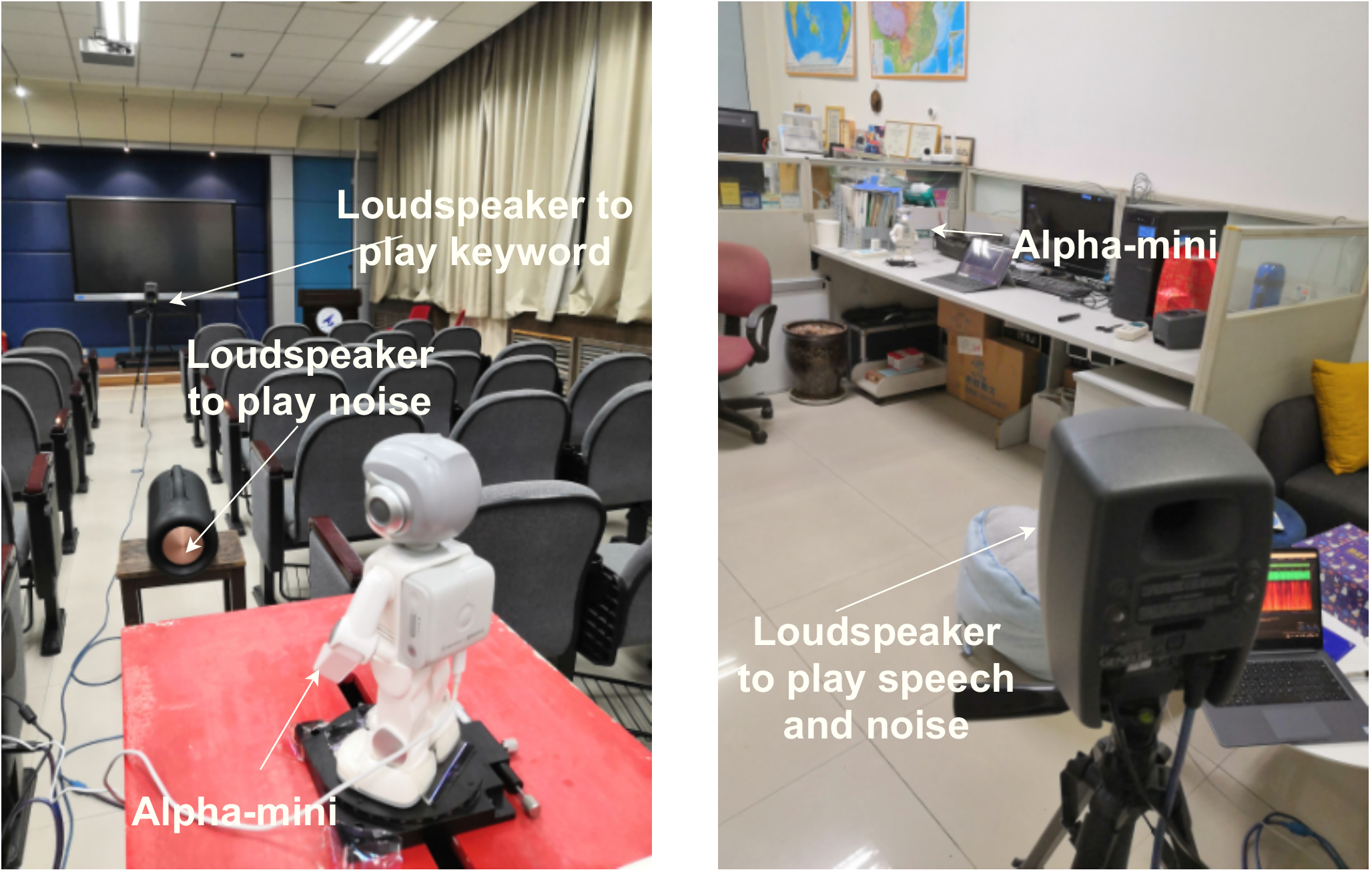}}   
  \caption{The typical recording scenes.}
  \label{fig:recording}
\end{figure}

\begin{figure}[!htb]
  \begin{minipage}[b]{1.0\linewidth}
    \centering
    \centerline{\includegraphics[width=4.6cm]{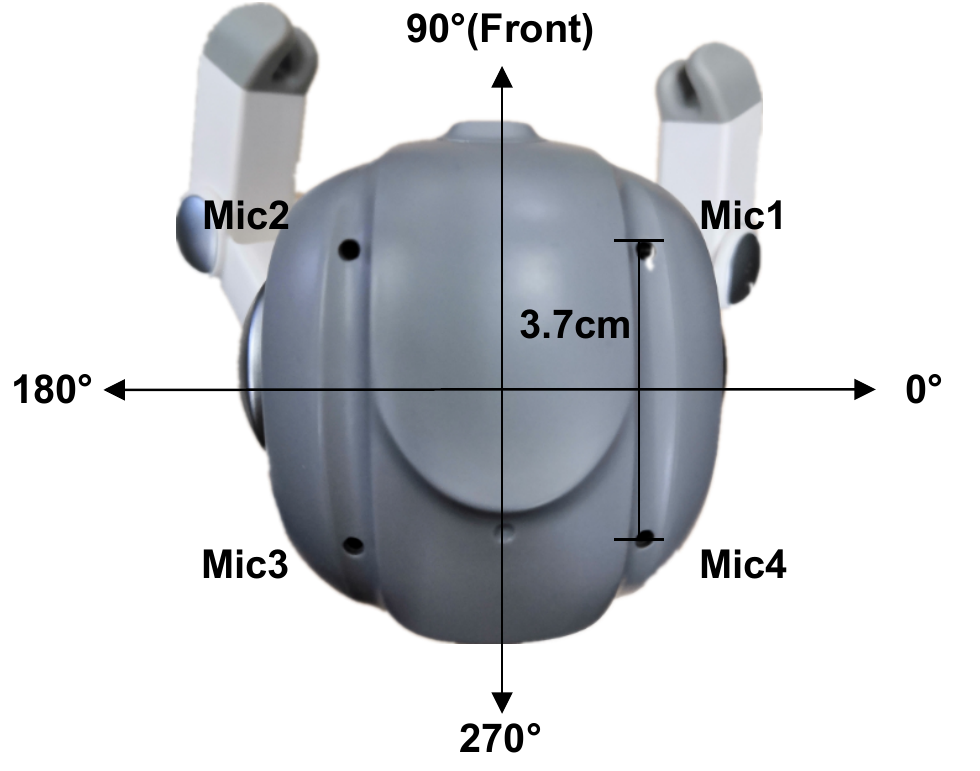}}
    \centerline{(a) Top view}\medskip
  \end{minipage}
  \begin{minipage}[b]{.48\linewidth}
    \centering
    \centerline{\includegraphics[width=3.8cm]{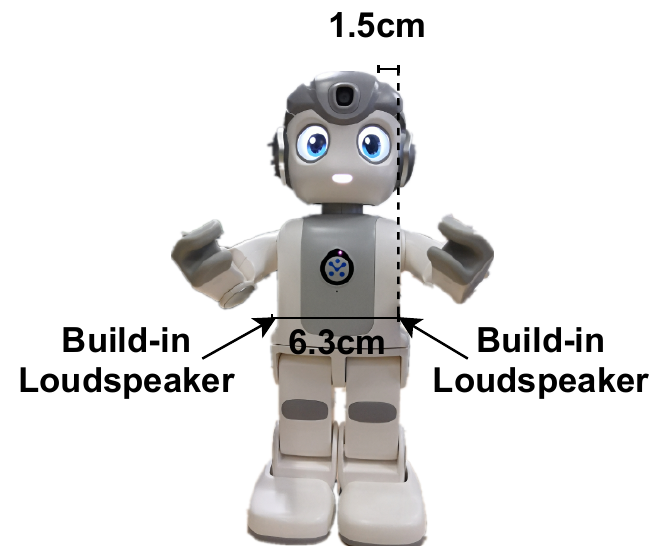}}
    \centerline{(b) Front view}\medskip
  \end{minipage}
  \hfill
  \begin{minipage}[b]{0.50\linewidth}
    \centering
    \centerline{\includegraphics[width=2.4cm]{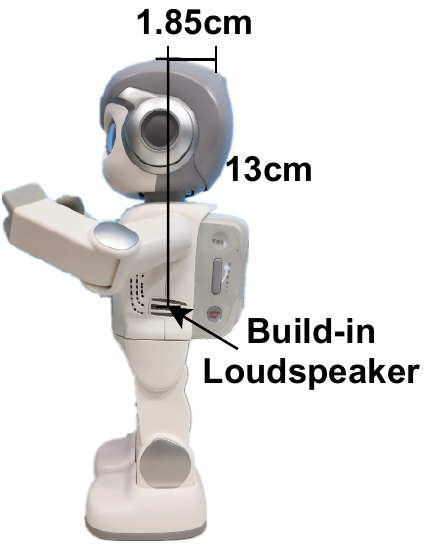}}
    \centerline{(c) Left view}\medskip
  \end{minipage}
  \caption{Three views of Alpha-mini robot.}
  \label{fig:threeviews}

    \end{figure}

Our goal of releasing the open source dataset in Table~\ref{table:subset} is to ensure the fair training resources and evaluation platform for researchers. The training data includes single channel keyword, speech, noise, echo data and recorded echo and mechanical noise of Alpha-mini. The development and evaluation sets contain keyword, speech, noise, echo and mechanical noise data recorded by Alpha-mini. During recording, we play the clean and noise signals through Hi-Fi loudspeakers and use the built-in microphone array of Alpha-mini to record. As for the echo data, various types of audio played by Alpha-mini built-in dual-loudspeaker is recorded by the built-in microphone array on the head of the Alpha-mini. The mechanical noise is generated by movable joints of Alpha-mini and recorded by the same built-in microphone array. Typical recording scenes are shown in Fig.~\ref{fig:recording}. The robot is equipped with four microphones located on its head and two loudspeakers located on both sides of its waist. The distance between two neighbor microphones and two loudspeakers are 3.7 cm and 6.3 cm, respectively. The vertical distance between the loudspeakers and the plane of microphone array is 13 cm, as shown in Fig.~\ref{fig:threeviews}. All recorded data are six-channel signal, where the first four channels are recorded signals and the rest two channels are reference signals played by the dual-loudspeakers of Alpha-mini. \par
\begin{table*}[!htb]
\caption{Data to release.}
\label{table:subset}
\centering
\scalebox{0.90}{
\begin{tabular}{cccccc}
\toprule
\multirow{2}{*}{Dataset}     & \multirow{2}{*}{Subset} & \multirow{2}{*}{Duration (hrs)} & \multirow{2}{*}{Format}                                                                      & \multirow{2}{*}{Scenario}                                                                                                     & \multirow{2}{*}{\begin{tabular}[c]{@{}c@{}}Mic-Loudspeaker \\ distance (metres)\end{tabular}} \\
                             &                         &                                 &                                                                                              &                                                                                                                               &                                                                                               \\ \midrule
\multirow{6}{*}{Training}    & Keyword-Train           & 9.4                            & \multirow{4}{*}{\begin{tabular}[c]{@{}c@{}}16kHz, 16bit, \\ single channel wav\end{tabular}} & \multirow{6}{*}{-}                                                                                                            & \multirow{6}{*}{-}                                                                            \\
                             & Speech-Train            & 146.1                           &                                                                                              &                                                                                                                               &                                                                                               \\
                             & Noise-Train             & 60.0                            &                                                                                              &                                                                                                                               &                                                                                               \\
                             & Echo-Train              & 28.5                            &                                                                                              &                                                                                                                               &                                                                                               \\ \cmidrule{2-4}
                             & Echo-Record             & 3.0                             & \multirow{2}{*}{\begin{tabular}[c]{@{}c@{}}16kHz, 16bit,\\ six-channel wav\end{tabular}}     &                                                                                                                               &                                                                                               \\
                             & Noise-Mech              & 8.6                             &                                                                                              &                                                                                                                               &                                                                                               \\ \midrule
\multirow{6}{*}{Development} & KWS-Dev                 & 7.5                             & \multirow{6}{*}{\begin{tabular}[c]{@{}c@{}}16kHz, 16bit,\\ six-channel wav\end{tabular}}     & \begin{tabular}[c]{@{}c@{}}Keyword only\\ Keyword+Noise\\ Keyword+Echo\\ Keyworkd+Noise+Echo\\ Keyword+Echo+Mech\end{tabular} & \multirow{6}{*}{{[}2, 4{]}}                                                                   \\ \cmidrule{2-3} \cmidrule{5-5}
                             & SSL-Dev                 & 20.0                            &                                                                                              & \begin{tabular}[c]{@{}c@{}}Speech only\\ Speech+Noise\\ Speech+Echo\\ Speech+Noise+Echo\\ Speech+Echo+Mech\end{tabular}      &                                                                                               \\ \midrule
\multirow{2}{*}{Evaluation}  & KWS-Eval                & \multirow{2}{*}{TBA}                            & \multirow{2}{*}{Same as Development}                                                         & \multirow{2}{*}{Same as Development}                                                                                          & \multirow{2}{*}{{[}2, 5{]}}                                                                   \\ \cmidrule{2-2}
                             & SSL-Eval                &                             &                                                                                              &                                                                                                                               &                                                                                               \\ \bottomrule
\end{tabular}
}
\end{table*}

Here we give the detailed illustration of subsets in Table~\ref{table:subset}:
 \begin{list}{\labelitemi}{\leftmargin=1em}
    \setlength{\topmargin}{0pt}
    \setlength{\itemsep}{0em}
    \setlength{\parskip}{0pt}
    \setlength{\parsep}{0pt}
  \item \textbf{Keyword-Train}:  `Wukong Wukong' wake-up word speech data provided by UBTECH recorded in anechoic room including voices from both adults and children, used for KWS model training.

  \item \textbf{Speech-Train}: The open-source AISHELL-1~\cite{bu2017aishell} training set is processed by deep complex convolutional recurrent network (DCCRN)~\cite{hu2020dccrn} and weighted prediction Error (WPE) algorithm~\cite{Nakatani2010Speech}, resulting in enhanced and dereverbed `clean' and `dry' version, which can be used for KWS model and SSL model training.

  \item \textbf{Noise-Train}: The noise data comes from 1) songs and pure music, 2) noise set of DNS challenge~\cite{reddy2020interspeech} and 3) various kinds of indoor noise including but not limited to clicking, keyboard, door opening/closing, fan, bubble noise, etc. This set can be used in KWS model and SSL model training.

  \item \textbf{Echo-Train}: The data released for echo simulation, which includes 1) songs, pure music, news broadcasting, people crosstalk and 2) speech generated by the Alpha-mini text-to-speech engine.

  \item \textbf{Echo-Record}: Various types of audio, played by Alpha-mini built-in dual-loudspeakers, recorded (echo) by the Alpha-mini built-in microphone array in a quiet room. The audio types played by Alpha-mini are the same as Echo-Train but the audio data has no overlap.

  \item \textbf{Noise-Mech}: Noise of mechanical movements generated by the movable joints of Alpha-mini, recorded by the Alpha-mini built-in microphone array in a quiet room.

  \item \textbf{KWS-Dev}: Recorded keywords, noise, echo and mechanical noise by Alpha-mini in two rooms. Keywords and noise are played by two Hi-Fi loudspeakers while echo is generated by Alpha-mini at the same time. The mechanical noise is recorded along then added to the recorded noisy and echoic signal. The audio types of played keyword, noise, echo and mechanical noise are the same as corresponding sets in training but audio data has no overlap. The Hi-Fi loudspeakers are randomly placed in each room. This set is used as development set for KWS model optimization.

  \item \textbf{SSL-Dev}: Recorded speech, noise, echo and mechanical noise by Alpha-mini in two rooms. Speech and noise are played by one Hi-Fi loudspeaker separately. Echo and mechanical noise are generated by Alpha-mini separately. Then we mix speech, noise, echo and mechanical noise together. The speech comes from enhanced and dereverbed `clean' and `dry' version of AISHELL-1 development and test set. The audio types of played noise, echo and mechanical noise are the same as corresponding sets in training but audio data has no overlap. The angle between Alpha-mini and loudspeakers covers every single degree from $1^{\circ}$ to $360^{\circ}$. The angle definition is illustrated in Fig.~\ref{fig:threeviews}(a) and the straight ahead of the robot is defined as $90^{\circ}$. This set is used as development set for SSL model optimization.

  \item \textbf{KWS-Test}: Recorded keywords, noise, echo and mechanical noise by Alpha-mini in three rooms. Other setups are the same as KWS-Dev. This is the evaluation set for KWS Track.

  \item \textbf{SSL-Test}: Recorded speech, noise, echo and mechanical noise by Alpha-mini in three rooms. Other setups are the same as SSL-Dev. This is the evaluation set for SSL Track.
\end{list}

\section{Keyword Spotting (KWS) Track}
This track is designed for KWS task. We illustrate the data arrangement, evaluation and ranking method, rules and baseline methods and results in this section.
\subsection{Data Arragement}
\label{sec:datacanuse}
The data can be used in this track is shown in Table~\ref{table:kwsdata}. Participants can use their own room impulse response (RIR), either collected or simulated, for data augmentation to train the KWS model. Furthermore, Echo-Record and Noise-Mech are provided as the reference of time-delay of echo and mechanical noise of Alpha-mini, respectively. Participants can also use these data sets during training. KWS-Dev, SSL-Dev, KWS-Eval, SSL-Eval are six-channel recorded data. Participants can use KWS-Dev and SSL-Dev directly without any simulation to optimize the model.
\subsection{Evaluation and Ranking}
We use a combination of false reject rate (FRR) and false alarm rate (FAR) on KWS-Eval and SSL-Eval respectively as the criterion of the KWS performance. Suppose the evaluation set has $N_{\text{key}}$ examples with keyword and $N_{\text{non-key}}$ examples without keyword, we define FRR and FAR as follows:
\begin{equation}
\begin{aligned}
\text{FRR} = \frac{N_{\text{FR}}}{N_{\text{key}}},\quad
\text{FAR} = \frac{N_{\text{FA}}}{N_{\text{non-key}}},
\end{aligned}
\label{eq:FRR}
\end{equation}
where $N_{\text{FR}}$ is the number of examples with keyword but the KWS system gives a negative decision and $N_{\text{FA}}$ is the number of examples without keyword but the KWS system gives a positive decision. The final score of KWS is defined as:
\begin{equation}
\begin{aligned}
\text{Score}^{\text{KWS}} = \text{FRR}+\text{FAR}.
\end{aligned}
\label{eq:KWSscore}
\end{equation}

FRR and FAR are calculated on all examples in KWS-Eval and SSL-Eval respectively and the final rank is $\text{Score}^{\text{KWS}}$ calculated by Eq.~(\ref{eq:KWSscore}). The system has lower $\text{Score}^{\text{KWS}}$ will be ranked higher.\par

KWS-Eval and SSL-Eval will not be released before organizers notify the participants about the results. Participants need to provide the organizers with a docker image of a runnable KWS system. The executable file in the image needs to receive the list of data in KWS-Eval and SSL-Eval and outputs the result of KWS. The output determines whether the sample contains keyword. If keyword exists, the sample is labeled as 1, and 0 otherwise. A detailed technical support of the usage and submission of docker will be provided later.
\subsection{Rules}
The use of any other data that is not provided by organizers (except for RIR) is strictly prohibited. Furthermore, it is not allowed to use KWS-Dev and SSL-Dev to train the KWS model.\par
There is no limitation on KWS model structure and model training technology used by participants. And the KWS model can have a maximum of 500 ms look ahead. To infer the current frame $T$ (in ms), the algorithm can access any number of past frames but only 500 ms of future frames ($T$ + 500 ms). In case there are submitted systems with the same score, the system with lower time delay will be given a higher ranking. \par
\begin{table}[!htb]
  \caption{Data arrangement for KWS Track.}
  \label{table:kwsdata}
  \centering
    \scalebox{0.90}{
  \begin{tabular}{ccc}
\toprule
Train         & Development                                                                & Evaluation                                                                   \\ \midrule
Keyword-Train & \multirow{6}{*}{\begin{tabular}[c]{@{}c@{}}KWS-Dev\\ SSL-Dev\end{tabular}} & \multirow{6}{*}{\begin{tabular}[c]{@{}c@{}}KWS-Eval\\ SSL-Eval\end{tabular}} \\
Speech-Train  &                                                                            &                                                                              \\
Noise-Train   &                                                                            &                                                                              \\
Echo-Train    &                                                                            &                                                                              \\
Echo-Record   &                                                                            &                                                                              \\
Noise-Mech    &                                                                            &                                                                              \\ \bottomrule
\end{tabular}
}
\end{table}

\subsection{Baseline}
\textbf{Front-end}: We use signal-based front-end for pre-processing. We apply frequency least mean square (FLMS) algorithm for acoustic echo cancellation (AEC) and delay and sum beamforming (DSBF) with SSL estimated by GCC-PHAT on multi-channel signal to generate single-channel signal as the input of KWS system. \par
\textbf{Deep KWS}: The first baseline is based on deep KWS model. After pre-processing, we extract 40-dimension mel-filterbanks feature by a window of 25ms with a shift of 10ms as the input of deep KWS. An 8-layer dilation time-delay neural network (TDNN)~\cite{Waibel1990Phoneme} is used as the KWS model shown in Fig.~\ref{fig:KWSSSL}(a). The kernel size of the first four layers is 5, and 3 for the rest. Dilation rate of these layers loops among $\left \{1, 2, 4, 8\right \}$. There is a batch normalization (BN) layer with rectified linear unit (ReLU) activation function between each TDNN layer. A fully connection layer (FC) is applied to map the output of TDNN into two categories -- keyword and filler. Softmax function is used to generate the posterior probability of both keyword and non-keyword. \par
We use post processing in~\cite{chen2014small} to generate the keyword confidence score from the posterior probabilities. The system will wake up if the confidence exceeds a predefined threshold. First, we smooth the raw posterior probabilities from the model over a fixed time window of size $w_{smooth}$. Suppose $p_{t}$ is the raw posterior probabilities of keyword at frame $t$, smoothing is done by:
\begin{equation}
p^{'}_{t} = \frac{1}{t-h_{smooth}+1} \sum_{k=h_{smooth}}^{t} p_{k},
\end{equation}
where $h_{smooth} = \text{max}(1,t-w_{smooth}+1)$ is the index of the first frame within the smooth window. Thus the confidence score at frame $t$ is the smoothed posterior $p^{'}_{t}$.\par
For data simulation, the RT60 of RIRs we generate ranges from 0.2 s to 0.8 s with image method. The room size ranges from 3 m $\times$ 3 m to 8 m $\times$ 8 m and the hight is maintained at 3 m. The mic-loudspeaker distance ranges from 1.5 m to 5 m. Both SNR and SER range from -5 dB to 10 dB. During training, cross entropy is used as the loss function. The batch size and the initial learning rate is set to 128 and 0.001, respectively. We train the model for 50 epochs with Adam optimizer using PyTorch. The result is shown in Table~\ref{table:kws result}. Note that the performance of the baseline KWS system decreases rapidly in noisy and echo scenarios. In particular, echo poses a much bigger challenge to the model than noise, possibly because the source of echo is closer to the microphone thus the SER is relatively low. Compared with noise scenario, the $\text{Score}^{\text{SSL}}$ in echo scenario decreases 0.16 on average. In addition, the overall performance of the KWS system is worse in the conference room scene due to larger reverberation and mic-loudspeaker distance.\par
\textbf{Keyword-filler}: We provide another baseline based on Kaldi Hi-Mia recipe~\footnote{\url{https://github.com/kaldi-asr/kaldi/tree/master/egs/hi_mia/w1}}. The acoustic model accepts the mel-filterbanks feature of front-end output as input and outputs the posterior probabilities of probability density function-identification (pdf-id). We extract 71-dimension mel-filterbanks feature by a window of 25 ms with a shift of 10 ms. A 6-layer dilation TDNN with ReLU activation function is used to get the time domain information. Then, a fully connected layer maps the high-dimensional representation to the posterior probabilities of pdf-id. The model is trained for 2 epochs with 512 batch size. The learning rate degradation algorithm is shown in Eq.~\ref{eq:kaldilr})
\begin{equation}
\begin{aligned}
lr_{j}\!=\!lr_{0} \!\times\! \text{exp}(\frac{j}{S-1} \text{log}(\frac{lr_{S-1}}{lr_{0}})) \quad & j\!=\!0,\!...,\!S-1,
\end{aligned}
\label{eq:kaldilr}
\end{equation}
where $S$ denotes the total step of training and $lr_{j}$ denotes the learning rate at step $j$.
As for language model, a decoding graph is used to calculate confidence score of keyword from the posterior probability of the acoustic model. The decoding graph only accepts the phonemes included in the keyword and computes the score of the keyword. The result is shown in Table~\ref{table:kws result}. It is proved again that echo, rather than noise, poses a greater impact on KWS result. Furthermore, KWS performance degrades to a great extend in larger reverberation and mic-loudspeaker distance scenario. A complete Kaldi based baseline script will be provided later.
\begin{table*}[!htb]
  \caption{Results of KWS baseline.}
  \label{table:kws result}
    \centering
  \scalebox{0.71}{
\begin{tabular}{ccccccccccccc}
\toprule
\multirow{3}{*}{Room}            & \multirow{3}{*}{Set}      & \multirow{3}{*}{Scenario} & \multicolumn{2}{c}{FRR}   & \multicolumn{2}{c}{Average}                     & \multirow{3}{*}{Set}      & \multirow{3}{*}{Scenario} & \multicolumn{2}{c}{FAR}   & \multicolumn{2}{c}{Average}                     \\ \cmidrule{4-7}\cmidrule{10-13}
                                 &                           &                           & \begin{tabular}[c]{@{}c@{}}Deep\\ KWS\end{tabular} & \begin{tabular}[c]{@{}c@{}}Keyword-\\ filler\end{tabular} & \begin{tabular}[c]{@{}c@{}}Deep\\ KWS\end{tabular} & \begin{tabular}[c]{@{}c@{}}Keyword-\\ filler\end{tabular}         &                           &                           & \begin{tabular}[c]{@{}c@{}}Deep\\ KWS\end{tabular} & \begin{tabular}[c]{@{}c@{}}Keyword-\\ filler\end{tabular} & \begin{tabular}[c]{@{}c@{}}Deep\\ KWS\end{tabular} & \begin{tabular}[c]{@{}c@{}}Keyword-\\ filler\end{tabular}        \\ \midrule
\multirow{5}{*}{Office}          & \multirow{10}{*}{KWS-Dev} & Keyword only              & 0.10     & 0.01           & \multirow{10}{*}{0.32} & \multirow{10}{*}{0.35} & \multirow{10}{*}{SSL-Dev} & Speech only               & 0.02     & 0.03           & \multirow{10}{*}{0.19} & \multirow{10}{*}{0.14} \\
                                 &                           & Keyword+Noise             & 0.16     & 0.08           &                        &                        &                           & Speech+Noise              & 0.14     & 0.08           &                        &                        \\
                                 &                           & Keyword+Echo              & 0.26     & 0.31           &                        &                        &                           & Speech+Echo               & 0.21     & 0.31           &                        &                        \\
                                 &                           & Keyword+Noise+Echo        & 0.41     & 0.37           &                        &                        &                           & Speech+Noise+Echo         & 0.23     & 0.24           &                        &                        \\
                                 &                           & Keyword+Echo+Mech         & 0.30     & 0.36           &                        &                        &                           & Speech+Echo+Mech          & 0.27     & 0.16           &                        &                        \\ \cmidrule{1-1} \cmidrule{3-5} \cmidrule{9-11}
\multirow{5}{*}{Conference Room} &                           & Keyword only              & 0.13     & 0.14           &                        &                        &                           & Speech only               & 0.03     & 0.02           &                        &                        \\
                                 &                           & Keyword+Noise             & 0.31     & 0.36           &                        &                        &                           & Speech+Noise              & 0.21     & 0.05           &                        &                        \\
                                 &                           & Keyword+Echo              & 0.42     & 0.57           &                        &                        &                           & Speech+Echo               & 0.24     & 0.20           &                        &                        \\
                                 &                           & Keyword+Noise+Echo        & 0.62     & 0.64           &                        &                        &                           & Speech+Noise+Echo         & 0.24     & 0.14           &                        &                        \\
                                 &                           & Keyword+Echo+Mech         & 0.52     & 0.68           &                        &                        &                           & Speech+Echo+Mech          & 0.29     & 0.15           &                        &                        \\ \bottomrule
\end{tabular}
}
\end{table*}
\begin{table*}[!htb]
  \caption{Results of SSL baseline.}
  \label{table:ssl result}
    \centering
  \scalebox{0.75}{
    \begin{tabular}{ccccccccccc}
      \toprule
      Set                       & Room                             & Scenario          & $\text{ACC}_{10}$ (\%) & Average (\%)            & $\text{ACC}_{7.5}$ (\%) & Average (\%)            & $\text{ACC}_{5}$ (\%) & Average (\%)            & $\text{MAE} (^{\circ})$ & Average ($^{\circ}$)    \\ \midrule
      \multirow{10}{*}{SSL-Dev} & \multirow{5}{*}{Office}          & Speech only       & 61.67                  & \multirow{10}{*}{33.69} & 45.49                   & \multirow{10}{*}{24.23} & 33.96                 & \multirow{10}{*}{17.78} & 9.80                    & \multirow{10}{*}{38.50} \\
      &                                  & Speech+Noise      & 46.25                  &                         & 33.40                   &                         & 23.68                 &                         & 22.17                   &                         \\
      &                                  & Speech+Echo       & 28.96                  &                         & 21.32                   &                         & 16.04                 &                         & 35.64                   &                         \\
      &                                  & Speech+Noise+Echo & 19.65                  &                         & 15.28                   &                         & 10.97                 &                         & 53.37                   &                         \\
      &                                  & Speech+Echo+Mech  & 19.72                  &                         & 14.93                   &                         & 10.97                 &                         & 53.88                   &                         \\\cmidrule{2-2} \cmidrule{3-4} \cmidrule{6-6} \cmidrule{8-8} \cmidrule{10-10}
      & \multirow{5}{*}{Conference Room} & Speech only       & 57.99                  &                         & 39.37                   &                         & 27.57                 &                         & 11.36                   &                         \\
      &                                  & Speech+Noise      & 46.60                  &                         & 33.19                   &                         & 24.65                 &                         & 20.23                   &                         \\
      &                                  & Speech+Echo       & 23.61                  &                         & 15.90                   &                         & 11.60                 &                         & 51.45                   &                         \\
      &                                  & Speech+Noise+Echo & 15.83                  &                         & 11.74                   &                         & 8.75                  &                         & 64.07                   &                         \\
      &                                  & Speech+Echo+Mech  & 16.67                  &                         & 11.67                   &                         & 9.58                  &                         & 63.06                   &                         \\ \bottomrule
    \end{tabular}
  }
\end{table*}
\begin{table}[!htb]
  \caption{Data arrangement for SSL Track.}
  \label{table:ssldata}
  \centering
  \scalebox{0.90}{
    \begin{tabular}{ccc}
    \toprule
    Train        & Development              & Evaluation                \\ \midrule
    Speech-Train & \multirow{5}{*}{SSL-Dev} & \multirow{5}{*}{SSL-Eval} \\
    Noise-Train  &                          &                           \\
    Echo-Train   &                          &                           \\
    Echo-Record  &                          &                           \\
    Noise-Mech   &                          &                           \\ \bottomrule
  \end{tabular}}
\end{table}
\section{Sound Source Location (SSL) Track}
This track is designed for SSL task. We illustrate the data arrangement, evaluation and ranking method, rules and baseline of SSL task in this section.
\subsection{Data}
The data that participants can use in this track is shown in Table~\ref{table:ssldata}. Participants can also use their own RIR, either collected or simulated, for data augmentation to train the SSL model. Furthermore, Echo-Record and Noise-Mech are provided as the reference of time-delay of echo and mechanical noise of Alpha-mini, respectively. Participants can also use these data sets during training. SSL-Dev and SSL-Eval are six-channel recorded data. Participants can use SSL-Dev directly without any simulation to optimize the model.
\subsection{Evaluation and Ranking}
We use a combination of Mean Absolute Error (MAE) and accuracy (ACC) as the criterion of the SSL performance. With the list of absolute errors of angle $\left \{e_{i}\right \}, i=1,...N$, where $N$ is the number of examples, we compute the MAE as:
\begin{equation}
\begin{aligned}
\text{MAE} = \frac{1}{N}\sum_{i=1}^{N}e_{i}.
\end{aligned}
\label{eq:MAE}
\end{equation}
ACC under different tolerances $\delta $ is defined as:
\begin{equation}
\text{ACC}_{\delta}= \frac{1}{N} \sum_{i=1}^{N}a_{i},
\quad
a_{i}=
\begin{cases}
1 & \text{if }e_{i} \leqslant \delta \\
0 & \text{otherwise}
\end{cases},
\end{equation}
The final score of SSL is defined as:
\begin{equation}
\begin{aligned}
&\text{Score}^{\text{SSL}}\\
&\!=\!(0.3 \! \times\! \text{ACC}_{10}\! +\! 0.35 \!\times\! \text{ACC}_{7.5} \!+\! 0.35 \!\times\! \text{ACC}_{5}) \\
&\!+\!(1 \!-\! \text{MAE} / \text{MAE}_{\text{baseline}} ).
\end{aligned}
\label{eq:SSL eval}
\end{equation}\par
The final rank is computed according to ACC under each tolerance and MAE of all examples in SSL-Eval by Eq.~(\ref{eq:SSL eval}). The $\text{MAE}_{\text{baseline}}$ of SSL-Eval will be released by organizers. The system with higher score will be ranked higher.\par
SSL-Eval will not be released before organizers notify the participants about the results. Participants need to provide organizers with a docker image of a runnable SSL system. The executable file in the image needs to receive the list of data in SSL-Eval and outputs the result of SSL. The output determines the direction of speech ranges from $1^{\circ}$ to $360^{\circ}$.  A detailed technical support of the usage and submission of docker will be provided later.
\subsection{Rules}
The use of any other data that is not provided by organizers (except for RIR) is strictly prohibited. Furthermore, it is not allowed to use SSL-Eval and Keyword-Train to train the SSL model. \par
There is no limitation on the system architecture, models, training techniques and time delays. However, we encourage participants to develop models with better performance and lower time delay. In case the submitted systems with the same score, the system with lower time delay will be given higher ranking.
\subsection{Baseline}
\begin{figure}[!htb]
  \begin{minipage}[b]{0.35\linewidth}
    \centering
    \centerline{\includegraphics[width=4.5cm]{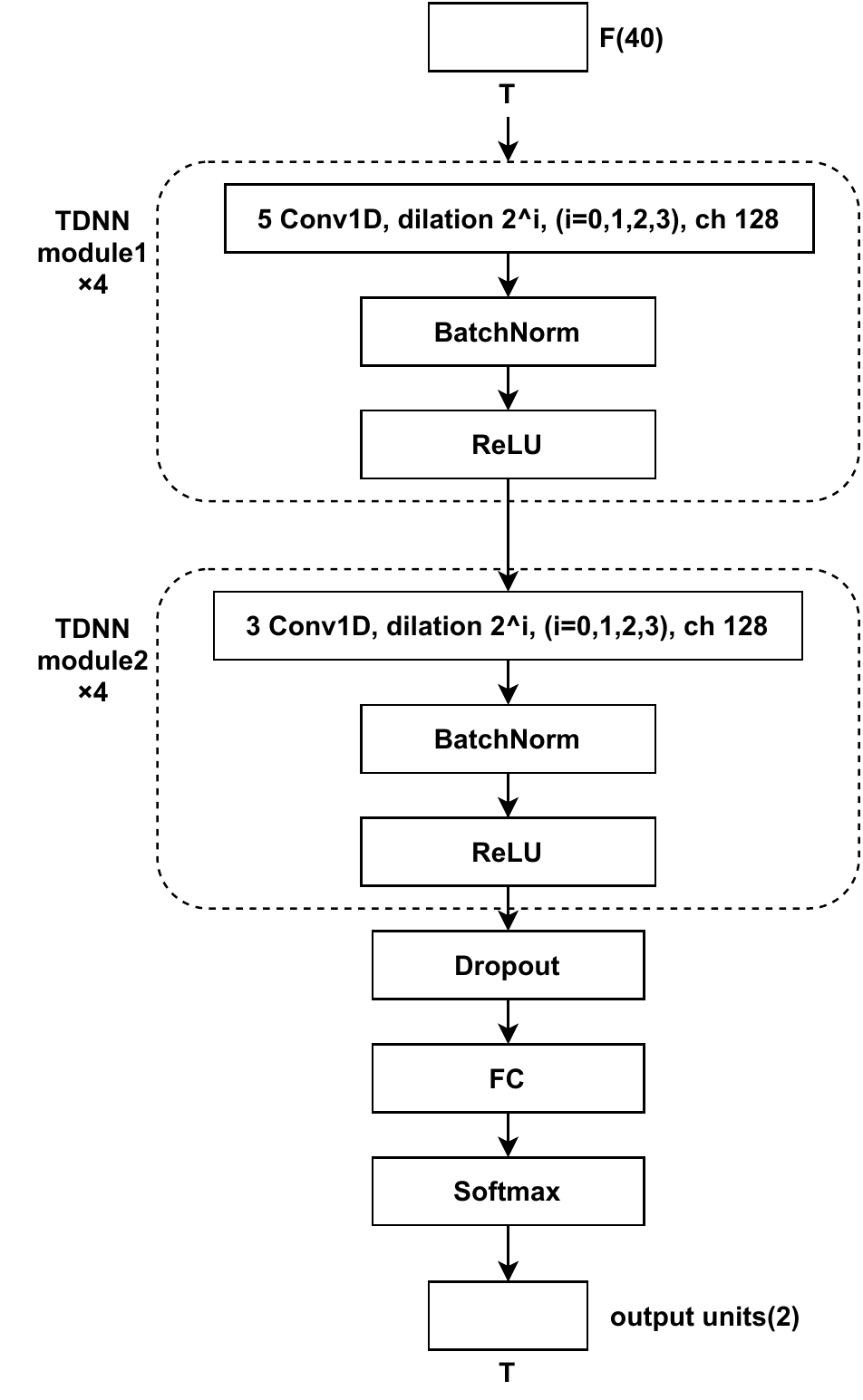}}
    \centerline{(a) Deep KWS baseline.}\medskip
  \end{minipage}
  \hfill
  \begin{minipage}[b]{0.80\linewidth}
    \centering
    \centerline{\includegraphics[width=4.6cm]{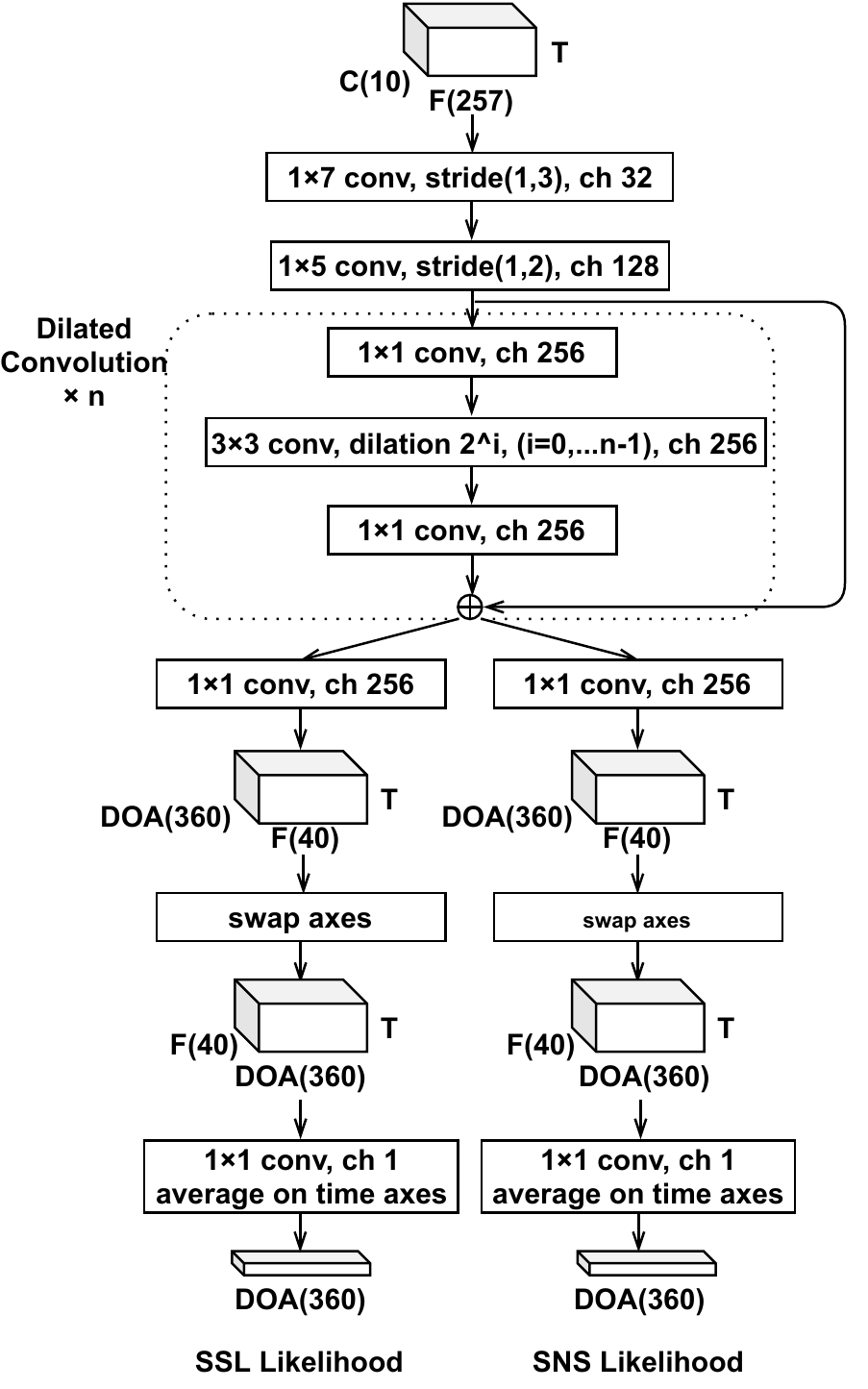}}
    \centerline{(b) SSL baseline.}\medskip
  \end{minipage}
  \caption{Model architecture of Deep KWS and SSL baseline.}
  \label{fig:KWSSSL}

\end{figure}

Inspired by~\cite{he2018deep,He2018Joint}, we adopt a fully convolutional multi-task framework for SSL task which takes multi-channel signal as input and output the probability distribution of the direction of sound source. We adapt short time Fourier transform (STFT) with 32ms frame length and 16ms frame hop to first five channels of the raw waveform and derive its magnitude and phase. Then we concatenate the magnitude and phase to generate $\mathbf{X} \in \mathbb{R}^{2C \times F\times T}$ as the input of the model, where $C$ denotes channel number, $F$ denotes the number of frequency bins, $T$ denotes the number of frames and 2 denotes magnitude and phase. The 3-layer temporal convolutional networks (TCN) module uses dilated convolution network whose dilation increase exponentially to get wider receptive field and more contextual information. The details of the model is shown in Fig.~\ref{fig:KWSSSL}(b). Multi-task is adopted to predict both the SSL and speech/non-speech (SNS) likelihood. The desired SSL output values are the maximum of Gaussian functions centered at the DOAs of the ground truth source:
\begin{equation}
p_{i}^{\text{SSL}}\!=\!
\begin{cases}
\text{max}_{\overline{\theta}\in \Theta }{\text{exp}(-d(\theta_{i},\overline{\theta})^{2}/\sigma ^{2})} & 1\!\leqslant\! i\!\leqslant\! 360 \\
0 & \text{otherwise}
\end{cases},
\end{equation}
where $\Theta = \Theta_{s} \cup \Theta_{n}$ is the union of ground truth speech and noise DOAs, $\sigma=45^{\circ}$ is the parameter to control the width of the Gaussian curves, $d(\cdot , \cdot)$ denotes the distance between two angles.
The desired SNS output values are the one-hot value depend on whether the nearest source is speech or noise:
\begin{equation}
p_{i}^{\text{SNS}}=
\begin{cases}
1 & \text{if the nearest source is speech} \\
0 & \text{otherwise}
\end{cases}.
\end{equation}
The loss function is defined as the mean square error (MSE) between estimated and ground truth SSL and SNS:
\begin{equation}
\text{Loss} = \left \| \mathbf{p}^{\text{SSL}}- \hat{\mathbf{p}}^{\text{SSL}}\right \|_{2}^{2} + \left \| \mathbf{p}^{\text{SNS}}- \hat{\mathbf{p}}^{\text{SNS}}\right \|_{2}^{2},
\end{equation}
where $\mathbf{p}^{\text{SSL}}, \hat{\mathbf{p}}^{\text{SSL}}, \mathbf{p}^{\text{SNS}}, \hat{\mathbf{p}}^{\text{SNS}} \in \mathbb{R}^{1\times360}$. During evaluation, the result of speaker location is defined by:
\begin{equation}
\hat{\theta} = \mathop{\text{argmax}} \limits_{1\leqslant i\leqslant 360}(\hat{\mathbf{p}}_{i}^{\text{SSL}} \cdot \hat{\mathbf{p}}_{i}^{\text{SNS}}).
\end{equation}\par
All hyper-parameters for data simulation are the same as KWS Track. We train the model for 20 epochs with Adam optimizer using PyTorch. Initial learning rate is set to 0.001 and will halve if no improvement on SSL-Dev. The result of baseline is shown in Table~\ref{table:ssl result}. It is worth noting that compared with the noise interference in far-field, the echo in near-field has a greater impact on the accuracy of SSL. Compared with Speech+Noise scenario, the ACC decreases 15.06 $\%$ and MAE increases 44.69$^{\circ}$ on average in Speech+Echo scenario. Furthermore, all results of Speech+Noise+Echo and Speech+Echo+Mech are very close, which indicates that mechanical noise also poses apparent impact on SSL accuracy.
\section{Important dates}
\label{sec:typestyle}
 \begin{list}{\labelitemi}{\leftmargin=1em}
    \setlength{\topmargin}{0pt}
    \setlength{\itemsep}{0em}
    \setlength{\parskip}{0pt}
    \setlength{\parsep}{0pt}
  \item September 27th, 2020: Registration due.
  \item September 30th, 2020: Release of the training and development set.
  \item November 22nd, 2020: Deadline for participants to submit docker mirror.
  \item December 6th, 2020: Organizers will notify the participants about the results.
  \item December 27th, 2020: Working note report deadline.
  \item January 19th-22nd, 2021: 2021 IEEE SLT Workshop date.
\end{list}
\section{Conclusions}
The IEEE SLT 2021 ASC is intended to promote research on KWS and SSL on humanoid robots in noise and echo scenarios. We provide train, development and evaluation datasets for participants to train and evaluate the model, as well as rules, evaluation methods and baselines as reference. It is expected that researchers from both academia and industry can advance the problem solving through this challenge.


\clearpage

\balance
\bibliographystyle{IEEEtran}
\bibliography{refs}

\begin{thebibliography}{10}
\providecommand{\url}[1]{#1}
\csname url@samestyle\endcsname
\providecommand{\newblock}{\relax}
\providecommand{\bibinfo}[2]{#2}
\providecommand{\BIBentrySTDinterwordspacing}{\spaceskip=0pt\relax}
\providecommand{\BIBentryALTinterwordstretchfactor}{4}
\providecommand{\BIBentryALTinterwordspacing}{\spaceskip=\fontdimen2\font plus
\BIBentryALTinterwordstretchfactor\fontdimen3\font minus
  \fontdimen4\font\relax}
\providecommand{\BIBforeignlanguage}[2]{{%
\expandafter\ifx\csname l@#1\endcsname\relax
\typeout{** WARNING: IEEEtran.bst: No hyphenation pattern has been}%
\typeout{** loaded for the language `#1'. Using the pattern for}%
\typeout{** the default language instead.}%
\else
\language=\csname l@#1\endcsname
\fi
#2}}
\providecommand{\BIBdecl}{\relax}
\BIBdecl

\bibitem{LVCSR}
J.~Mamou, B.~Ramabhadran, and O.~Siohan, ``Vocabulary independent spoken term
  detection,'' in \emph{Proceedings of the 30th annual international ACM SIGIR
  conference on Research and development in information retrieval}, 2007, pp.
  615--622.

\bibitem{LVCSR2}
P.~Motlicek, F.~Valente, and I.~Szoke, ``Improving acoustic based keyword
  spotting using {LVCSR} lattices,'' in \emph{2012 IEEE International
  Conference on Acoustics, Speech and Signal Processing (ICASSP)}.\hskip 1em
  plus 0.5em minus 0.4em\relax IEEE, 2012, pp. 4413--4416.

\bibitem{LVCSR3}
I.-F. Chen, C.~Ni, B.~P. Lim, N.~F. Chen, and C.-H. Lee, ``A novel keyword+
  {LVCSR}-filler based grammar network representation for spoken keyword
  search,'' in \emph{The 9th International Symposium on Chinese Spoken Language
  Processing}.\hskip 1em plus 0.5em minus 0.4em\relax IEEE, 2014, pp. 192--196.

\bibitem{HMM}
B.~Yan, R.~Guo, X.~Zhu, and B.~Zhang, ``An approach of keyword spotting based
  on {HMM},'' in \emph{Proceedings of the 3rd World Congress on Intelligent
  Control and Automation (Cat. No. 00EX393)}, vol.~4.\hskip 1em plus 0.5em
  minus 0.4em\relax IEEE, 2000, pp. 2757--2759.

\bibitem{HMM2}
J.~R. Rohlicek, W.~Russell, S.~Roukos, and H.~Gish, ``Continuous hidden
  {M}arkov modeling for speaker-independent word spotting,'' in
  \emph{International Conference on Acoustics, Speech, and Signal
  Processing,}.\hskip 1em plus 0.5em minus 0.4em\relax IEEE, 1989, pp.
  627--630.

\bibitem{Choisy2007Dynamic}
C.~Choisy, ``Dynamic handwritten keyword spotting based on the {NSHP-HMM},'' in
  \emph{Ninth International Conference on Document Analysis and Recognition
  (ICDAR 2007)}, vol.~1.\hskip 1em plus 0.5em minus 0.4em\relax IEEE, 2007, pp.
  242--246.

\bibitem{Hou2017Investigating}
J.~Hou, L.~Xie, and Z.~Fu, ``Investigating neural network based
  query-by-example keyword spotting approach for personalized wake-up word
  detection in {M}andarin {C}hinese,'' in \emph{ISCSLP}.\hskip 1em plus 0.5em
  minus 0.4em\relax IEEE, 2016, pp. 1--5.

\bibitem{Chen2015Query}
G.~Chen, C.~Parada, and T.~N. Sainath, ``Query-by-example keyword spotting
  using long short-term memory networks,'' in \emph{ICASSP}.\hskip 1em plus
  0.5em minus 0.4em\relax IEEE, 2015, pp. 5236--5240.

\bibitem{hou2020mining}
J.~Hou, Y.~Shi, M.~Ostendorf, M.-Y. Hwang, and L.~Xie, ``Mining effective
  negative training samples for keyword spotting,'' in \emph{ICASSP 2020-2020
  IEEE International Conference on Acoustics, Speech and Signal Processing
  (ICASSP)}.\hskip 1em plus 0.5em minus 0.4em\relax IEEE, 2020, pp. 7444--7448.

\bibitem{yuan2019verifying}
Y.~Yuan, Z.~Lv, S.~Huang, and L.~Xie, ``Verifying deep keyword spotting
  detection with acoustic word embeddings,'' in \emph{2019 IEEE Automatic
  Speech Recognition and Understanding Workshop (ASRU)}.\hskip 1em plus 0.5em
  minus 0.4em\relax IEEE, 2019, pp. 613--620.

\bibitem{yuan2020fast}
Y.~Yuan, L.~Xie, C.-C. Leung, H.~Chen, and B.~Ma, ``Fast query-by-example
  speech search using attention-based deep binary embeddings,'' \emph{IEEE/ACM
  Transactions on Audio, Speech, and Language Processing}, 2020.

\bibitem{DNN}
G.~Chen, C.~Parada, and G.~Heigold, ``Small-footprint keyword spotting using
  deep neural networks,'' in \emph{2014 IEEE International Conference on
  Acoustics, Speech and Signal Processing (ICASSP)}.\hskip 1em plus 0.5em minus
  0.4em\relax IEEE, 2014, pp. 4087--4091.

\bibitem{DNN2}
Z.~Chen, Y.~Qian, and K.~Yu, ``Sequence discriminative training for deep
  learning based acoustic keyword spotting,'' \emph{Speech Communication}, vol.
  102, pp. 100--111, 2018.

\bibitem{Retsinas2018Exploring}
G.~Retsinas, G.~Sfikas, N.~Stamatopoulos, G.~Louloudis, and B.~Gatos,
  ``Exploring critical aspects of {CNN}-based keyword spotting. a {PHOCN}et
  study,'' in \emph{IAPR}, 2018, pp. 13--18.

\bibitem{fernandez2007application}
S.~Fern{\'a}ndez, A.~Graves, and J.~Schmidhuber, ``An application of recurrent
  neural networks to discriminative keyword spotting,'' in \emph{International
  Conference on Artificial Neural Networks}.\hskip 1em plus 0.5em minus
  0.4em\relax Springer, 2007, pp. 220--229.

\bibitem{higuchi2020stacked}
T.~Higuchi, M.~Ghasemzadeh, K.~You, and C.~Dhir, ``Stacked 1{D} convolutional
  networks for end-to-end small footprint voice trigger detection,''
  \emph{arXiv preprint arXiv:2008.03405}, 2020.

\bibitem{adya2020hybrid}
S.~Adya, V.~Garg, S.~Sigtia, P.~Simha, and C.~Dhir, ``Hybrid
  {T}ransformer/{CTC} networks for hardware efficient voice triggering,''
  \emph{arXiv preprint arXiv:2008.02323}, 2020.

\bibitem{wang2019virtual}
X.~Wang, S.~Sun, and L.~Xie, ``Virtual adversarial training for ds-cnn based
  small-footprint keyword spotting,'' in \emph{2019 IEEE Automatic Speech
  Recognition and Understanding Workshop (ASRU)}.\hskip 1em plus 0.5em minus
  0.4em\relax IEEE, 2019, pp. 607--612.

\bibitem{Shan2018}
C.~Shan, J.~Zhang, Y.~Wang, and L.~Xie, ``Attention-based end-to-end models for
  small-footprint keyword spotting,'' \emph{arXiv preprint arXiv:1803.10916},
  2018.

\bibitem{wu2018monophone}
M.~Wu, S.~Panchapagesan, M.~Sun, J.~Gu, R.~Thomas, S.~N.~P. Vitaladevuni,
  B.~Hoffmeister, and A.~Mandal, ``Monophone-based background modeling for
  two-stage on-device wake word detection,'' in \emph{2018 IEEE International
  Conference on Acoustics, Speech and Signal Processing (ICASSP)}.\hskip 1em
  plus 0.5em minus 0.4em\relax IEEE, 2018, pp. 5494--5498.

\bibitem{sigtia2020multi}
S.~Sigtia, E.~Marchi, S.~Kajarekar, D.~Naik, and J.~Bridle, ``Multi-task
  learning for speaker verification and voice trigger detection,'' in
  \emph{ICASSP 2020-2020 IEEE International Conference on Acoustics, Speech and
  Signal Processing (ICASSP)}.\hskip 1em plus 0.5em minus 0.4em\relax IEEE,
  2020, pp. 6844--6848.

\bibitem{knapp1976generalized}
C.~Knapp and G.~Carter, ``The generalized correlation method for estimation of
  time delay,'' \emph{IEEE transactions on acoustics, speech, and signal
  processing}, vol.~24, no.~4, pp. 320--327, 1976.

\bibitem{Schmidt1986Multiple}
R.~Schmidt, ``Multiple emitter location and signal parameter estimation,''
  \emph{IEEE transactions on antennas and propagation}, vol.~34, no.~3, pp.
  276--280, 1986.

\bibitem{Lin2018Reverberation}
S.~Lin, ``Reverberation-robust localization of speakers using distinct speech
  onsets and multichannel cross correlations,'' \emph{IEEE/ACM Transactions on
  Audio, Speech, and Language Processing}, vol.~26, no.~11, pp. 2098--2111,
  2018.

\bibitem{Lin2018JOINTLY}
------, ``Jointly tracking and separating speech sources using multiple
  features and the generalized labeled multi-bernoulli framework,'' in
  \emph{IEEE International Conference on Acoustics, Speech and Signal
  Processing (ICASSP)}, 2018, pp. 3211--3215.

\bibitem{pertila2017robust}
P.~Pertil{\"a} and E.~Cakir, ``Robust direction estimation with convolutional
  neural networks based steered response power,'' in \emph{2017 IEEE
  International Conference on Acoustics, Speech and Signal Processing
  (ICASSP)}.\hskip 1em plus 0.5em minus 0.4em\relax IEEE, 2017, pp. 6125--6129.

\bibitem{Xu2017Weighted}
C.~Xu, X.~Xiong, S.~Sun, R.~Wei, and H.~Li, ``Weighted spatial covariance
  matrix estimation for music based {TDOA} estimation of speech source,'' in
  \emph{Interspeech}, 2017, pp. 1894--1898.

\bibitem{Wang2018Robust}
Z.~Wang, X.~Zhang, and D.~Wang, ``Robust speaker localization guided by deep
  learning-based time-frequency masking,'' \emph{IEEE/ACM Transactions on
  Audio, Speech, and Language Processing}, vol.~27, no.~1, pp. 178--188, 2018.

\bibitem{he2018deep}
W.~He, P.~Motlicek, and J.-M. Odobez, ``Deep neural networks for multiple
  speaker detection and localization,'' in \emph{2018 IEEE International
  Conference on Robotics and Automation (ICRA)}.\hskip 1em plus 0.5em minus
  0.4em\relax IEEE, 2018, pp. 74--79.

\bibitem{Lollmann2018The}
H.~W. L{\"o}llmann, C.~Evers, A.~Schmidt, H.~Mellmann, and W.~Kellermann, ``The
  {LOCATA} challenge data corpus for acoustic source localization and
  tracking,'' in \emph{2018 IEEE 10th Sensor Array and Multichannel Signal
  Processing Workshop (SAM)}, 2018, pp. 410--414.

\bibitem{bu2017aishell}
H.~Bu, J.~Du, X.~Na, B.~Wu, and H.~Zheng, ``{Aishell}-1: An open-source
  mandarin speech corpus and a speech recognition baseline,'' in \emph{2017
  20th Conference of the Oriental Chapter of the International Coordinating
  Committee on Speech Databases and Speech I/O Systems and Assessment
  (O-COCOSDA)}.\hskip 1em plus 0.5em minus 0.4em\relax IEEE, 2017, pp. 1--5.

\bibitem{hu2020dccrn}
Y.~Hu, Y.~Liu, S.~Lv, M.~Xing, S.~Zhang, Y.~Fu, J.~Wu, B.~Zhang, and L.~Xie,
  ``{DCCRN}: Deep complex convolution recurrent network for phase-aware speech
  enhancement,'' \emph{arXiv preprint arXiv:2008.00264}, 2020.

\bibitem{Nakatani2010Speech}
T.~Nakatani, T.~Yoshioka, K.~Kinoshita, M.~Miyoshi, and B.~H. Juang, ``Speech
  dereverberation based on variance-normalized delayed linear prediction,''
  \emph{IEEE Transactions on Audio Speech Language Processing}, vol.~18, no.~7,
  pp. 1717--1731, 2010.

\bibitem{reddy2020interspeech}
C.~K. Reddy, E.~Beyrami, H.~Dubey, V.~Gopal, R.~Cheng, R.~Cutler,
  S.~Matusevych, R.~Aichner, A.~Aazami, S.~Braun \emph{et~al.}, ``The
  {i}nterspeech 2020 deep noise suppression challenge: Datasets, subjective
  speech quality and testing framework,'' \emph{arXiv preprint
  arXiv:2001.08662}, 2020.

\bibitem{Waibel1990Phoneme}
A.~Waibel, T.~Hanazawa, G.~E. Hinton, K.~Shikano, and K.~J. Lang, ``Phoneme
  recognition using time-delay neural networks,'' \emph{Readings in Speech
  Recognition}, vol.~1, no.~3, pp. 393--404, 1990.

\bibitem{chen2014small}
G.~Chen, C.~Parada, and G.~Heigold, ``Small-footprint keyword spotting using
  deep neural networks,'' in \emph{2014 IEEE International Conference on
  Acoustics, Speech and Signal Processing (ICASSP)}.\hskip 1em plus 0.5em minus
  0.4em\relax IEEE, 2014, pp. 4087--4091.

\bibitem{He2018Joint}
W.~He, P.~Motlicek, and J.-M. Odobez, ``Joint localization and classification
  of multiple sound sources using a multi-task neural network,'' in
  \emph{Interspeech 2018}, 2018.

\end{thebibliography}

\end{document}